\documentclass{article}[12pt]
\usepackage[utf8]{inputenc}
\usepackage{amsmath}
\usepackage{amssymb}
\usepackage{lipsum}
\usepackage{bm}
\usepackage{graphicx}
\usepackage{graphics}
\usepackage[dvipdf]{epsfig}
\usepackage{float}
\usepackage{wrapfig}
\usepackage{url}
\usepackage{epstopdf} 
\usepackage{xcolor}
\usepackage[margin=1in]{geometry}
\usepackage{amsmath}
\usepackage{amsbsy}
\usepackage{amssymb}
\usepackage{amscd}
\usepackage{amsfonts}

\newcommand{\beq}{\begin{equation}}
\newcommand{\eeq}{\end{equation}}
\newcommand{\beqs}{\begin{eqnarray}}
\newcommand{\eeqs}{\end{eqnarray}}
\newcommand{\beql}{\begin{equation} \label}

\newcommand{\bfzero}{\mathbf{0}}

\newcommand{\grad}{\mathop{\rm grad}\nolimits}
\newcommand{\divergence}{\mathop{\rm div}\nolimits}
\newcommand{\curl}{\mathop{\rm curl}\nolimits}

\begin{document}

\title{Field Dislocation Mechanics and Phase Field Crystal models}
\author{Amit Acharya\\Department of Civil and Environmental     Engineering,\\ and Center for Nonlinear Analysis,\\
Carnegie Mellon University,\\
 Pittsburgh, Pennsylvania 15213, USA
\and Jorge Vi\~nals\\ School of Physics and Astronomy, University of Minnesota,\\ Minneapolis, Minnesota 55455, USA}
\date{\today}

\maketitle
\begin{abstract}
\noindent A new formulation of the Phase Field Crystal model is presented that is consistent with the necessary microscopic independence between the phase field, reflecting the broken symmetry of the phase, and both mass density and elastic distortion. Although these quantities are related in equilibrium through a macroscopic equation of state, they are independent variables in the free energy, and can be independently varied in evaluating the dissipation functional that leads to the model governing equations. The equations obtained describe dislocation motion in an elastically stressed solid, and serve as an extension of the equations of plasticity to the Phase Field Crystal setting. Both finite and small deformation theories are considered, and the corresponding kinetic equations for the fields derived.
\end{abstract}

\section{Introduction}

\noindent The Phase Field Crystal (PFC) model was introduced as a mesoscale description of a nonequilibrium crystalline phase, valid at the molecular length scale, but only over long, diffusive time scales  \cite{re:elder02}. By eliminating the need to resolve the time scale associated with lattice vibration, the Phase Field Crystal model has become a widely used computational tool capable of describing a wide variety of phenomena in materials science \cite{re:emmerich12}. One of the strengths of the formulation is the ease in the description of defected solids, including, for example, dislocation dissociation, stacking fault formation, grain boundary motion, and coarsening of polycrystalline configurations. Further spatial coarse graining has also been undertaken, leading to models in which the characteristic spatial variation is also slow compared with the molecular length scale \cite{re:goldenfeld05,re:elder10,re:yeon10,re:praetorius19}. 

The model begins with the introduction of a phenomenological, non convex free energy functional, $\Phi_{sh}[\psi]$, of a phase field $\psi(\mathbf{x},t)$ and its gradients. Although we will not explicitly use this functional form below, we mention the widely used form
\begin{equation}
    \Phi_{sh}[\psi] = \int_{\Omega} d \bm{x} \; \varphi_{sh} = \int_{\Omega} d \bm{x} \; \left[ \frac{1}{2} \left[ (\nabla^{2} + q_{0}^{2}) \psi \right]^{2} - \frac{\epsilon}{2} \psi^{2} + \frac{1}{4}\psi^{4} \right],
\end{equation}
where in this dimensionless units $q_{0} = 1$ (we retain the notation $q_{0}$ for ease of discussion below), and $0 < \epsilon \ll 1$ is the dimensionless control parameter of the bifurcation between the ground states $\psi = 0$ and $\psi$ periodic. We also introduce $\overline{\psi}$, the conserved spatial average of $\psi$, as a control parameter. The combination of gradients is chosen so as to produce ground states that are spatially periodic, with characteristic wavenumber $q_{0}$. The choice of nonlinearity and the value of $\overline{\psi}$ determine the symmetry of the resulting ground state lattice. While the bulk of the early work focused on two dimensional hexagonal lattices, research has also considered three dimensional systems, including fcc and bcc lattices \cite{re:elder10}, and specific materials such as, for example, Fe \cite{re:pisutha-arnond13} or graphene \cite{re:huter16}. We will assume below that $\Phi_{sh}[\psi]$ is given for a three dimensional system, but will not focus on its specific properties which have been extensively studied elsewhere (e.g., in Ref. \cite{re:huter16}).

Phase Field Crystal model free energies have been derived by using Density Functional Theory methods, with the expectation of obtaining functionals that capture the long time diffusive evolution of the mass density as the relevant order parameter \cite{re:elder07,re:huang10}. The free energies obtained provide a reasonable description of the freezing phase transition \cite{re:archer19}. However, extensions to include the momentum density in the set of slow or hydrodynamic variables have not been considered to the same extent, except for colloidal systems \cite{re:archer09}, and, more recently, in the so called hydrodynamic formulation of the Phase Field Crystal \cite{re:heinonen16}. In this latter case, both mass and momentum conservation are considered at the mesoscale (still involving spatial variations that are slow compared with the lattice spacing $1/q_{0}$). For weak distortions around the ground state, a smooth displacement field can be introduced, resulting in a dynamical dispersion relation that includes both phonon propagation and damping, in agreement with standard theory. Notably, the dispersion at large wavenumber becomes entirely diffusive as diffusion of the phase field controls the local relaxation of a weakly distorted configuration. Although the study does not address how to explicitly incorporate topological constraints necessary to describe a defected configuration, results are given for grain rotation and shrinkage in a two dimensional, hexagonal phase. Grain radius is seen to decay with time as $t^{-1/2}$, as expected. The amplitude of the decay rate increases with increasing Newtonian viscosity in the momentum equation. In the limit of large viscosity, the results of the overdamped model of Ref. \cite{re:heinonen14} for the grain size as a function of time are recovered. Since the boundary of the grain comprises a periodic array of dislocations, this example indicates that the theory is capable of describing the evolution of an initially defected configuration.

However, the Phase Field Crystal model has some important shortcomings that point to its incompleteness. In most research to date, the mass density and lattice distortion of the crystalline phase are generally described by the same scalar field $\psi$. If this is the case, then their variations are not independent. Consider, for example, that $\psi$ is a conserved mass density. Then, its local variation through distortion is $\delta \psi = - \partial_{k} (\psi \delta u_{k})$, where $u_{k}$ is the $k$-th component of the displacement vector, the phase of $\psi$. From this relation, the variation $\delta \Phi_{sh}/\delta u_{k} = \psi \partial_{k} (\delta \Phi_{sh}/\delta \psi)$. Since the stress is defined through $\partial_{j} T_{ij} = - \delta \Phi_{sh}/\delta u_{i}$, then $\partial_{j} T_{ij} = - \psi \partial_{i}(\delta \Phi_{sh}/\delta \psi)$.
%As a consequence,
%$$
%\divergence \frac{\delta \Phi_{sh}}{\delta \grad \bm{u}} = - \psi \grad \frac{\delta \Phi_{sh}}{\delta \psi}
%$$
%{\color{blue} (can you please provide the details of this result - I would like to %understand it - it is too cryptic for me as written. thanks.)}
%where $\bm{u}$ is a small displacement from the ground state, $\psi(\bm{x}') = %\psi(\bm{x}+\bm{u}(\bm{x}))$. 
%
This relation is correct in equilibrium where both sides of the equation vanish, but not in general outside of equilibrium. Furthermore, if both variations are not considered to be independent, then lattice distortions can only relax diffusively, which is unphysical. This difficulty has been recognized for a long time, and a number of modified models have been introduced to allow for relaxation of the phase field in a time scale faster than diffusion \cite{re:stefanovic06,re:majaniemi07,re:heinonen14,re:zhou19}, including the hydrodynamic formulation alluded to above \cite{re:heinonen16}.

Despite modifications to the Phase Field Crystal model in order to accelerate the relaxation of elastic distortions, restricting the model to a single field $\psi$ still leads to difficulties or inconsistencies. One such difficulty involves the definition of physical system boundaries, and the imposition of boundary conditions involving domain shape or traction. The specification of boundary traction, for example, needs to be done indirectly through manipulation of the phase field. In their study of the motion of a single dislocation under an imposed strain, Berry et al. \cite{re:berry06} rigidly displaced a small layer of sites at the boundary. The resulting distortion propagated into the bulk system slowly (diffusively), thus preventing direct control of the stress field in the defect region other than readjusting the displacement of the boundary layer, and waiting for a long time until the bulk stress would readjust. The ensuing motion of the dislocation is quite different from would be expected from classical elasticity and the Peach-K\"{o}hler force \cite{re:skaugen18b,re:salvalaglio20}. A second issue concerns the recent result that the ground state of the Phase Field Crystal appears to be, in fact, under a large pressure. For example, for the model parameters that are employed to describe bcc Fe, the ground state pressure is as large as $1.8 \times 10^{6}$ atm at melting \cite{re:pisutha-arnond13}. Whether this state of pressure is or is not taken into account in the determination of the linear elastic constants from the phase field free energy, it is possible to predict both a decrease or an increase in their values as a function of $\overline{\psi}$ (related to average density or pressure) \cite{re:wang18}. The proper definition of strain from the phase field has been further discussed in Ref. \cite{re:huter16} which suggests holding the value of $\overline{\psi}$ constant under volume change, which implies that it is not related to the mass density. Finally, modeling plastic motion of defects within the Phase Field Crystal leads to another class of difficulties. Elastic and plastic distortions are independent, and ordinarily relax over widely different time scales. While it is well understood that mass and lattice defect velocities are independent quantities \cite{re:kosevich79}, they are simultaneously described by a single scalar quantity $\psi$ in the Phase Field Crystal model. 

The approach that we propose here is based on the realization that the PFC/Swift-Hohenberg functional does not posses intrinsic elasticity; indeed, the Swift-Hohenberg functional, despite its elegance and immense generality, contains no information on the forces that hold matter together either based on Quantum Mechanics or macroscopic elastic response, and this is borne out by the fitting it requires, see, e.g., \cite[Eqn. (65)]{re:huter16}. Therefore, we use it as a mathematical device or indicator function that (i) describes the symmetry of a crystalline lattice even when locally deformed, (ii) serves to locate topological defects and provide for their topological index, and, (iii) allows to  conserve topological charge in close to equilibrium processes involving defect motion through its `phase' being constrained to equal a field (described below) whose mechanics explicitly satisfies a conservation law for (signed) topological charge allowing defect nucleation and annihilation. We introduce a configurational distortion tensor $\bm{P}$, a pointwise functional of the phase field $\psi$, which coincides with the inverse elastic distortion tensor of the medium $\bm{W}$ only in equilibrium. Away from equilibrium, we allow relative fluctuations between both such that the elastic response is captured by $\bm{W}$, and the diffusive relaxation by $\bm{P}$. Section \ref{sec:theory} describes our theory in general for nonlinear distortions, whereas Sec. \ref{sec:theory-linear} considers the approximation of small elastic distortions so as to compare our results with existing models.

The fully nonlinear (geometric and material) dynamics of the inverse elastic distortion field is governed by the partial differential equation based model of Field Dislocation Mechanics (FDM) \cite{acharya2001model, acharya2004constitutive, acharya2006size,
acharya2011microcanonical, arora2020dislocation, arora2020finite,arora2020unification}. It completes the program of the theory of continuously distributed dislocations \cite[and earlier references therein]{kazuo1963non}, \cite{bilby1955continuous,kroener1971continuum,kroner1981continuum,mura1963continuous,fox1966continuum,willis1967second,re:kosevich79} extended from its origins in linear elasticity and links between differential geometry and defect kinematics to a full-fledged nonlinear theory of continuum mechanics accounting for equations of balance, evolution, large irreversible material deformations (plasticity), material inertia and dissipation, geometric and material nonlinearity in finite bodies of arbitrary elastic anisotropy subjected to general boundary and initial conditions, and understood at a level of granularity suitable for computer implementation to obtain approximate solutions \cite{arora2020finite,zhang2015single}, \cite[and following works for the geometrically linear model]{roy2005finite}. FDM is `fluid-like' in its description of the behavior of solids with defects in not relying on the existence of a reference configuration or a plastic distortion tensor, while predicting physically observed large, irreversible plastic deformation of the body due to the motion of dislocations (as well as recoverable elastic deformation and residual stress). The coupled FDM-PFC model we propose shares all of these important properties.

In closing this Introduction, we mention the Phase Field models of dislocations \cite{wang2001nanoscale,koslowski2002phase,shen2003phase, rodney2003phase,denoual2004dynamic,re:levitas12,re:mianroodi15,mianroodi2016theoretical} that have been quite successful in solving a variety of problems related to dislocation mechanics close to equilibrium. These models are restricted to small deformation kinematics and the notion of plastic strains from a fixed reference configuration (that is not physically determinable from an internally stressed defected initial state). More importantly, Phase Field models require the definition of the so-called `crystalline energy' or the `Generalized Stacking Fault energy' that has to be defined creatively from the a-priori knowledge of slip-systems of a material and an atomistic $\gamma$-surface procedure first introduced by Vitek in \cite{vitek1968intrinsic}. As a consequence, the number of independent fields included in the model is related to the number of slip systems identified and considered \cite{rodney2003phase,re:levitas12}, and dislocation combination rules need to be adapted accordingly \cite{shen2003phase}. This is different from PFC which \emph{predicts} material symmetry and consequent defect motions on preferred planes and directions dictated by that symmetry \cite{re:yamanaka17}.  Furthermore, the dynamics of phase field models rely on an Allen-Cahn gradient flow for a set of non-conserved scalar `disorder' fields (or non-convex incremental energy minimization with highly non-unique solutions as in \cite{koslowski2002phase}), with one consequence being that a spatially homogeneous phase field can evolve based on the levels of stress and energy density fields. This is in contrast to FDM where evolution of the elastic distortion (beyond `convection') can only occur at a field point where a dislocation exists (i.e., the curl of the distortion does not vanish), regardless of the level of stress or energy density at that point. This `thermodynamic driving force' property follows from the second law of thermodynamics constrained by an explicit condition of conservation of Burgers vector (topological charge) during the evolution of elastic distortion - and is a feature that is consistent with the form of the Peach-K\"{o}hler force of classical dislocation theory.

\section{Finite deformation phase field crystal theory of dislocation motion}
\label{sec:theory}

\subsection{Choice of fields}
\label{sec:variables}

We focus on an isothermal system and consider a  simply-connected body (even in the presence of line defects) at all times. The following set of independent variables is introduced: $\rho$, the continuum mass density, the material velocity $\bm{v}$, $\bm{W}$, the inverse elastic distortion, and $\psi$ the phase field. The tensor field $\bm{W}$ maps the (linear approximation to the) deformed elastic lattice pointwise to the undeformed lattice (the latter assumed known). In the absence of line defects, $\curl \bm{W} = \bfzero$ (compatible elasticity), a potential field $\bm{X}$ defining a reference configuration exists in which the undeformed lattice can be embedded: $d X_{i} = \frac{\partial X_{i}}{\partial x_{j}} dx_{j} = F^{-1}_{ij} dx_{j}$, with $\bm{F}^{-1} = \bm{W}$. In terms of a displacement field $\bm{u}$ of the reference (which exists in the compatible case), the tensor $U_{ij} = \partial_{i} u_{j} = \partial_{i}(x_{j} - X_{j}) = \delta_{ij} - F^{-1}_{ij}$, so that $\bm{W} = \bm{F}^{-1} = \bm{I} - \bm{U}$. Even in the incompatible case, defining $\bm{W}^{-1} - \bm{I} = \bm{U}$ and assuming $|\bm{U}| \ll 1$, $\bm{W} \approx \bm{I} - \bm{U}$.

The key ingredient of our model is a new (two-point) second rank tensor $\bm{P}$ (standing for \emph{phase}) with the same symmetry properties under rotation as $\bm{W}$. Its value at each point in the material is a functional of the phase field $\psi$, and is defined so as to describe the distortion of the surfaces of constant $\psi$. After averaging the phase field over a scale on the order of $q_{0}^{-1}$ \cite{re:skaugen18}, one can define a triad of local wavevectors $\bm{q}^{n}$, different than those of the ground state of $\Phi_{sh}[\psi]$, the latter denoted by $\bm{q}_{0}^{n}$. Then we define  $\bm{q}_{0}^{n} = \bm{P}^{-T} \bm{q}^{n}$.
%
% Transpose or inverse - check
%
The tensor $\bm{P}$ describes a local configurational distortion that can be associated with the field $\psi$, without endowing the phase field with any elastic properties. 
%%Note that although $\bm{W}$ is defined as a gradient of the displacement field, its curl can be non zero at dislocations (an incompatible displacement field).
Note that the curl of the tensor field $\bm{W} \neq \bfzero$ in general, and $\bm{P}$ will not vanish at defects in the phase field equivalent lattice.

\subsection{Balance equations}

The density $\rho$ satisfies mass conservation
\begin{equation}
    \dot{\rho} + \rho \divergence \bm{v} = 0
    \label{eq:mass_con}
\end{equation}
where $\dot{(~)}$ represents a material time derivative, and $\bm{v}$ is the material velocity (center of mass velocity of an element of volume), and all spatial differential operators at any given time are on the configuration occupied by the body at that time. Momentum conservation is written as
\begin{equation}
    \rho \dot{\bm{v}} = \divergence \bm{T} + \rho \bm{b}
    \label{eq:mom_con}
\end{equation}
where $\bm{T}$ is the stress tensor, which in the present context, is symmetric, and $\bm{b}$ is a specified body force density (per unit mass). For quasi-static motions of the body, we simply write $ \divergence \bm{T} + \rho \bm{b} = \bfzero$.

If the medium contains dislocation lines, the inverse elastic distortion is imcompatible, and we write \cite{willis1967second}
\begin{equation}
    \curl \bm{W} = \curl \bm{P} = - \bm{\alpha},
    \label{eq:alpha_def}
\end{equation}
where $\bm{\alpha}$ is the dislocation density tensor. The integral of this tensor over a surface equals the sum of the Burgers' vectors of the dislocation lines that thread the surface. Motion of the dislocation lines induces a change in the distortion tensor given by \cite{acharya2004constitutive,acharya2011microcanonical}
\begin{equation}
    \dot{\bm{W}} + \bm{W} \bm{L} = \bm{\alpha} \times \bm{V}
    \label{eq:Vdef}
\end{equation}
where we have introduced the tensor $\bm{L} = \grad \bm{v}$, and the local dislocation line velocity $\bm{V}$ relative to the local mass velocity. This equation is implied by topological charge conservation under defect motion (up to a gradient of a vector field that can be assumed to vanish for microscopic defect motions) \cite{acharya2015dislocation} and, conversely, enforces such conservation when operative.

\subsection{Free energy, dissipation inequality, and governing equations}

We next consider the free energy density of the system $\varphi$ to be a function not only of $\rho$, $\bm{W}$ and $\psi$, but also of $\bm{P}$, treated as an independent variable,
\begin{eqnarray}
\int_\Omega d \bm{x} \; \rho \varphi(\rho,\bm{W},\psi,\bm{P}) = \int_\Omega d\bm{x} \; \rho \varphi_{e}(\rho,\bm{W},\bm{P}) & + & C_{sh} \Phi_{sh}[\psi] + \nonumber \\ + \frac{C_{w}}{2} \int_\Omega d \bm{x} \; \rho | \bm{W} - \bm{P} |^{2} & + & \frac{C_{\rho}}{2} \int_\Omega d \bm{x} \; \rho (\rho-\psi)^{2}.
\label{eq:fe}
\end{eqnarray}
The first term in the right hand side of Eq. (\ref{eq:fe}) is the standard elastic energy. We allow a dependence on $\bm{P}$ only to express the fact that the actual functional form of the elastic constant matrix will depend on the symmetry of the lattice, and that potentially on the linear elastic constants that will themselves depend on that symmetry, and the local state of distortion of the phase field. For the simplest extension of linear elasticity to rotationally invariant nonlinear elasticity, for example, one would write
\begin{equation}
    \varphi_{e} = \frac{1}{2 \rho_{0}} \bm{E}:\bm{C}(\bm{P}):\bm{E},
\end{equation}
where $\bm{C}$ is the tensor of elastic moduli, possibly dependent on $\bm{P}$, and $\bm{E}$ is the symmetric strain tensor $\bm{E} = \frac{1}{2} \left( \bm{F}^{eT}\bm{F}^e - \bm{I} \right)$, with $\bm{F}^e := \bm{W}^{-1}$. 

For simplicity, we introduce the notation
\begin{equation}
    \Phi_{wp} = \int_\Omega d\bm{x} \; \rho \varphi_{e}(\rho,\bm{W},\bm{P}) + \frac{C_{w}}{2} \int_\Omega d \bm{x} \; \rho | \bm{W} - \bm{P} |^{2},
    \label{eq:phiwp}
\end{equation}
which is also, implicitly, a functional of the phase field $\psi$ (through $\bm{P}$). The coupling constants $C_{sh}, C_{w}$ and $C_{\rho}$ are non negative, and we will typically focus on the case in which $C_{w}$ is large $C_{w} \gg | \bm{C} |$. 

Motivated by Eq. \eqref{eq:phiwp} and the evolution of $\bm{P}$ necessary for response due to a superposed rigid motion on a given motion of a body in which $\psi$ does not evolve, we assume that 
\begin{equation}
    \int_\Omega d\bm{x} \; \frac{\delta \Phi_{wp}}{\delta \psi} \dot{\psi} = \int_\Omega d\bm{x} \; \rho \frac{\partial}{\partial \bm{P}} \left( \varphi_{e} + C_{w} \varphi_{wp} \right) : \left[ \dot{\bm{P}} + \bm{P} \bm{L} \right].
    \label{eq:sh_diss}
\end{equation}
where we have defined $\varphi_{wp} = \frac{1}{2} |\bm{W}-\bm{P}|^{2}$.
%%This relation explicitly accounts for the independence between variations in the phase field configuration (in $\psi$) and variations in the density $\rho$ and lattice distortion $\bm{W}$.

With the explicit form of the conservation laws, and the form of the free energy introduced, we can use a dissipation inequality to derive the kinetic laws governing the evolution of the fields. We write the Second Law of Thermodynamics in the form
\begin{equation}
    \int_{\partial \Omega} \left( \bm{T} \cdot \hat{\bm n} \right) \cdot \bm{v} dS + \int_{\Omega} d \bm{x}\, \rho \bm{b} \ge \frac{d}{dt} \int_{\Omega} d \bm{x} \; \rho \varphi + \frac{d}{dt} \int_{\Omega} d \bm{x} \; \frac{1}{2} \rho |\bm{v}|^{2},
\end{equation}
so that the power expended by external agencies (applied traction on the outer boundary and the applied body forces) is greater or equal to the rate of change of the free energy plus kinetic energy. Integrating this relation by parts and using the balance of linear momentum, we write
\begin{equation}
    \int_{\Omega} d \bm{x} \; \bm{T}:\bm{L} - \frac{d}{dt} \int_{\Omega} d \bm{x} \; \rho \varphi \ge 0.
     \label{eq:second_law}
\end{equation}

By explicit substitution of Eq. (\ref{eq:fe}), one finds
\begin{eqnarray}
     \int_{\Omega} d \bm{x} \; \bm{T}:\bm{L} & - & \int_{\Omega} d \bm{x} \; \rho \left( \frac{\partial \varphi_{e}}{\partial \bm{W}} + C_{w} \frac{\partial \varphi_{wp}}{\partial \bm{W}}\right) : (- \bm{W} \bm{L} + \bm{\alpha}\times\bm{V}) \nonumber \\
     &-& \int_{\Omega} d \bm{x}\; \left[ \varphi_{e} + C_{w} \varphi_{wp} + C_{\rho}\varphi_{\rho} + C_{\rho} (\rho - \psi) \right] \left( - \rho {\rm Tr}(\bm{L}) \right) -  \label{eq:second_law_2} \\
     &-& \int_{\Omega} d\bm{x} \; \left[ C_{sh} \frac{\delta \Phi_{sh}}{\delta \psi} +C_{\rho} \rho (\rho - \psi) \right] \dot{\psi} - \int_{\Omega} d\bm{x} \; \rho \left[ \frac{\partial \varphi_{e}}{\partial \bm{P}} + C_{w} \frac{\partial \varphi_{wp}}{\partial \bm{P}} \right] : \dot{\bm{P}} \ge 0 \nonumber
\end{eqnarray}
By using Eq. (\ref{eq:sh_diss}), the last term in the L.H.S. of Eq. (\ref{eq:second_law_2}) can be written as
$$
- \int_{\Omega} d\bm{x} \; \rho \left[ \frac{\partial \varphi_{e}}{\partial \bm{P}} + C_{w} \frac{\partial \varphi_{wp}}{\partial \bm{P}} \right] : (-\bm{P} \bm{L}) - \int_{\Omega} d \bm{x} \; \frac{\delta \Phi_{wp}}{\delta \psi} \dot{\psi}
$$

This equation can be further rewritten to highlight products of thermodynamics forces and currents as
\begin{eqnarray}
& & \int_{\Omega} d \bm{x} \; \left[  \bm{T} + \rho \bm{W}^{T}  \left( \frac{\partial \varphi_{e}}{\partial \bm{W}} + C_{w} \frac{\partial \varphi_{wp}}{\partial \bm{W}} \right) + \rho a 
 \bm{I} \right] : \bm{L} \nonumber \\
 & - & \int_{\Omega} d \bm{x} \; \rho \left( \frac{\partial \varphi_{e}}{\partial \bm{W}} + C_{w} \frac{\partial \varphi_{wp}}{\partial \bm{W}} \right) : (\bm{\alpha}\times \bm{V}) \nonumber \\
 & + & \int_{\Omega} d \bm{x}\; \rho \bm{P}^{T} \left( \frac{\partial \varphi_{e}}{\partial \bm{P}} + C_{w} \frac{\partial \varphi_{wp}}{\partial \bm{P}} \right) : \bm{L} \label{eq:second_law_3} \\
 & - & \int_{\Omega} d \bm{x}\; \left[ C_{sh}\frac{\delta \Phi_{sh}}{\delta \psi} + C_{\rho} \rho(\rho - \psi) + \frac{\delta \Phi_{wp}}{\delta \psi} \right] \dot{\psi} \ge 0 \nonumber
\end{eqnarray}
where we have defined $a = \varphi_{e} + C_{w} \varphi_{wp} + C_{\rho} \varphi_{\rho} + C_{\rho} (\rho - \psi)$.

This expression can be further simplified since the free energy density $\varphi$ is invariant under rotation. In that case, the antisymmetric (or skew) part
$$
\left( \bm{W}^{T} \frac{\partial \varphi}{\partial \bm{W}} + \bm{P}^{T} \frac{\partial \varphi}{\partial \bm{P}} \right)_{\rm skew} = \bfzero.
$$
Therefore of the terms proportional to $\bm{L}$ in Eq. (\ref{eq:second_law_3}), only those proportional to the symmetric part of velocity gradient $\bm{D} = (\bm{L}+\bm{L}^{T})/2$ contribute. We combine them into
\begin{equation} 
\int_{\Omega} d \bm{x}\; \left\{ \bm{T} + \rho \left[ \bm{W}^{T} \left( \frac{\partial \varphi_{e}}{\partial \bm{W}} + C_{w} \frac{\varphi_{wp}}{\partial \bm{W}} \right) + \bm{P}^{T} \left( \frac{\partial \varphi_{e}}{\partial \bm{P}} + C_{w} \frac{\varphi_{wp}}{\partial \bm{P}} \right) + a \bm{I} \right] \right\} : \bm{D} 
\label{eq:second_law_4}
\end{equation}

This completes our calculation of the dissipation inequality. One can now identify the reversible parts of the various currents, followed by the introduction of the respective dissipative currents in order to respect the inequality. The symmetric reversible stress follows directly from Eq. (\ref{eq:second_law_4}),
\begin{equation}
    \bm{T}^{R} = - \rho \left[ \bm{W}^{T} \left( \frac{\partial \varphi_{e}}{\partial \bm{W}} + C_{w} \frac{\varphi_{wp}}{\partial \bm{W}} \right) + \bm{P}^{T} \left( \frac{\partial \varphi_{e}}{\partial \bm{P}} + C_{w} \frac{\varphi_{wp}}{\partial \bm{P}} \right) + a \bm{I} \right]
    \label{eq:stress_r}
\end{equation}
Since our formulation applies not only to crystalline phases, but also to other phases with broken symmetries still described by a phase field, we mention that it is possible to introduce a dissipative stress as $\bm{T}^{D} = \bm{\eta}:\bm{D}$, where $\bm{\eta}$ is a fourth rank viscosity tensor. The number of independent components of the elastic constant and viscosity tensors depend the the symmetry of the system, and have been enumerated for several important cases \cite{re:martin72}.

We will restrict our analysis to dissipative defect velocities only. In order to ensure positivity of dissipation, we write
\begin{equation}
\bm{V} = - \bm{M} \, \bm{X} : \left[ \rho \left( \frac{\partial \varphi_{e}}{\partial \bm{W}} + C_{w} \frac{\partial \varphi_{wp}}{\partial \bm{W}} \right)^{T} \bm{\alpha}\right]
\label{eq:VD_def}
\end{equation}
where $\bm{M}$ is a positive definite mobility tensor. For $C_w = 0$, it can be shown that the driving force in the above relation corresponds to the exact generalization of the form of the Peach-K\"{o}hler force to the fully nonlinear setting \cite{acharya2004constitutive}.

Finally, we identify the reversible and irreversible currents of the phase field $\psi$. The condition for reversible motion is simply $\dot{\psi} = 0$, that is, advection of the phase field. The dissipative component is chosen to enforce positivity, leading to an order parameter equation,
\begin{equation}
    \dot{\psi} = - L \left[  C_{sh}\frac{\delta \Phi_{sh}}{\delta \psi} + C_{\rho} \rho(\rho - \psi) + \frac{\delta \Phi_{wp}}{\delta \psi} \right]
    \label{eq:op_evol}
\end{equation}
where the constant $L > 0$ is the phase field mobility. Importantly, although mass is a conserved quantity, the phase field that describes the broken symmetry is not. On this particular, our model does differ from implementations of the Phase Crystal model based on density functional theory in which the order parameter is chosen to be the mass density.

In summary, the complete set of equations includes mass (Eq. (\ref{eq:mass_con})), momentum (Eq. (\ref{eq:mom_con})), and topological charge (Eq. (\ref{eq:Vdef}) conservation, along with the definition of the dislocation tensor, Eq. (\ref{eq:alpha_def}). The phenomenological currents that follow from the dissipation inequality and the model free energy, Eq. \eqref{eq:fe}, are the stress, Eq. (\ref{eq:stress_r}), the defect velocity Eq. (\ref{eq:VD_def}), and the evolution equation for the phase field, Eq. (\ref{eq:op_evol}).

Before considering the small deformation limit of the model, we outline several qualitative features of the evolution of a defected phase as given by the governing equations. An initially defected configuration will be described by an order parameter field $\psi$. Topological defects will be located in regions of non zero curl of $\bm{P}$, with $\bm{P}$ defined by a point wise oriented triad in reciprocal space from $\psi$ \cite{re:skaugen18b}, compared to the same object for the ground state of $\Phi_{sh}$. For $ C_{w}, C_{sh}$ large and of  comparable magnitude, the order parameter will relax quickly (and diffusively) to a local minimum of 
$$
C_{sh} \Psi_{sh} + \frac{ C_{w}}{2} \int d \bm{x} \; \rho |\bm{W}-\bm{P}|^{2}
$$
relatively independently of the resulting changes induced in the elastic energy $\varphi_{e}$, and in mass density fluctuations. This process will be accompanied by the relaxation of the elastic distortion in phonon lifetime scales, also quickly if the quasistatic elastic limit is invoked. Further evolution will be slow, driven by the Peach-K\"{o}hler force in Eq. (\ref{eq:VD_def}), which is dominated by the elastic stress term $\partial \varphi_{e}/\partial \bm{W}$. If the configuration is not defected, but subjected to body forces, traction and/or velocity boundary conditions, the solution of the elasticity problem will yield $\bm{W}$, which will - if $C_{w}$ and $C_{sh}$ are large - quickly modify $\psi$. In this case of no defects, $\psi$ becomes a passive indicator function mediating nonlinear anisotropic elastic response up to homogeneous nucleation of defects.

\section{Small deformation limit}
\label{sec:theory-linear}

In the small deformation or geometrically linear limit, we consider a fixed simply connected reference configuration for the body and assume that the deforming body remains close to this configuration at all times so that all spatial derivatives can be written w.r.t. this fixed reference configuration. As is customary, it is also formally assumed that various distortion measures are `small' in magnitude. In this case, as mentioned in Sec. \ref{sec:variables}, the inverse elastic distortion is $\bm{W} = \bm{I} - \bm{U}$ and we treat $\bm{U}$ as the fundamental measure of elastic distortion. We note that $\curl \bm{U} \neq \bfzero$ in the presence of defects, when it cannot be written as a gradient of a displacement field. We will also consider the symmetrized elastic distortion $\bm{\epsilon} = \bm{U}_{sym}$, $\epsilon_{ij} = (1/2)(U_{ij}+U_{ji})$. Analogously, we define $\bm{Q} = \bm{I} - \bm{P}$. 
%
% I think the metric g_{kl} = \delta_{kl} + 2\epsilon_{kl}
%

From Eqs. (\ref{eq:alpha_def}) and (\ref{eq:Vdef}), the equations defining the dislocation density tensor and defect motion are now 
\begin{equation}
    \curl \bm{U} = \bm{\alpha}, \quad\quad \bm{L} = \dot{\bm{U}} + \bm{\alpha} \times \bm{V}
    \label{eq:kinematics_sd}
\end{equation}
where we have neglected the quadratic term $\bm{U}\bm{L}$. These equations are the classical equations of plastic motion \cite{re:kosevich79, mura1963continuous}. Here, $\bm{L}$ is still the velocity gradient, but now with respect to the fixed reference configuration.

In analogy to Eq. (\ref{eq:fe}) we write the free energy density as
\begin{eqnarray}
\int_{\Omega} d \bm{x} \; \varphi(\rho,\bm{U},\psi,\bm{Q}) = \int_{\Omega} d\bm{x} \; \varphi_{e}(\rho,\bm{U},\bm{Q}) & + & C_{sh} \Phi_{sh}[\psi] + \nonumber \\ + \frac{C_{w}}{2} \int_{\Omega} d \bm{x} \; | \bm{U} - \bm{Q} |^{2} & + & \frac{C_{\rho}}{2} \int_{\Omega} d \bm{x} \; (\rho-\psi)^{2}.
\label{eq:fesd}
\end{eqnarray}

In the small deformation regime, the dissipation inequality is written as
\begin{equation}
    \int_{\Omega} d \bm{x}\; \bm{T} : \bm{L} -  \int_{\Omega} d \bm{x}\; \dot{\varphi} \ge 0.
    \label{eq:second_lawsd}
\end{equation}
As in Sec. \ref{sec:theory}, we define
\begin{equation}
    \Phi_{uq} = \int_{\Omega} d \bm{x}\; \varphi_{e}(\bm{U},\bm{Q}) + \frac{C_{w}}{2} \int_{\Omega} d \bm{x}\; |\bm{U}-\bm{Q}|^{2}
\end{equation}
The second term of Eq. (\ref{eq:second_lawsd}) can now be written as
$$
    \int_{\Omega} d \bm{x}\; \dot{\varphi} =  \int_{\Omega} d \bm{x}\; \frac{\partial \varphi_{e}}{\partial \bm{U}} : (\bm{L} - \bm{\alpha}\times\bm{V}) +  \int_{\Omega} d \bm{x}\; \frac{\delta \Phi_{uq}}{\delta \psi} \dot{\psi} + C_{sh} \int_{\Omega} d \bm{x}\; \frac{\delta \Phi_{sh}}{\delta \psi} \dot{\psi} 
$$
\begin{equation}
+ C_{\rho} \int_{\Omega} d \bm{x}\; \left( \rho - \psi \right) \left(-\rho {\rm Tr} ( \bm{L}) \right) - C_{\rho} \int_{\Omega} d \bm{x}\; (\rho - \psi)\dot{\psi}, 
    \label{eq:omegadot_sd}
\end{equation}
where we have used the relation, analogous to Eq.( \ref{eq:sh_diss}),
\begin{equation}
    \int_{\Omega} d \bm{x}\; \frac{\delta \Phi_{uq}}{\delta \psi} \dot{\psi} = \int_{\Omega} d \bm{x}\; \left[ \frac{\partial \varphi_{e}}{\partial \bm{Q}} + C_{w} \frac{\partial \varphi_{uq}}{\partial \bm{Q}} \right] : \dot{\bm{Q}}.
    \label{eq:dissip2}
\end{equation}
Complete invariance properties under superposed rigid motions is not customarily considered in the geometrically linear theory and hence certain nonlinear terms like $\bm{Q}\bm{L}$ in (\ref{eq:dissip2}) do not appear in Eq. (\ref{eq:omegadot_sd}).

Since the stress tensor is symmetric, and (infinitesimal) rotational invariance requires that the dependence of $\varphi_{e}$ on $\bm{U}$ be only through the symmetrized distortion $\bm{\epsilon}$, the dissipation relation Eq. (\ref{eq:second_lawsd}) can be written as,
$$
\int_{\Omega} d \bm{x} \left[ \bm{T} - \frac{\partial \varphi_{e}}{\partial \bm{\epsilon}} + C_{\rho} \rho (\rho - \psi) \bm{I} \right] : \bm{L}_{sym} + \int_{\Omega} d \bm{x}\; \frac{\partial \varphi_{e}}{\partial \bm{\epsilon}} : (\bm{\alpha} \times \bm{V}) + 
$$
\begin{equation}
    + \int_{\Omega} \left[ - \frac{\delta \Phi_{uq}}{\delta \psi} - C_{sh} \frac{\delta \Phi_{sh}}{\delta \psi} + C_{\rho} \frac{\Phi_{\rho\psi}}{\delta \psi} \right] \dot{\psi} \ge 0,
\end{equation}
where we have used the notation
$$
\Phi_{\rho \psi}= \frac{1}{2} \int_{\Omega} d \bm{x}\; (\rho - \psi)^{2}.
$$

With this form of the dissipation inequality, we can identify the stress and the remaining quantities. The reversible part of the stress is
\begin{equation}
    \bm{T}^{R} = \frac{\partial \varphi_{e}}{\partial \bm{\epsilon}} - C_{\rho} \rho(\rho-\psi) \bm{I},
    \label{eq:stress_sd}
\end{equation}
with the dissipative part nominally given by the same expression as in Sec. \ref{sec:theory}. The defect velocity is the standard Peach-K\"{o}hler force,
\begin{equation}
    \bm{V} = \bm{M} \bm{X} :\left[ \left( \frac{\partial \varphi_{e}}{\partial \bm{\epsilon}} \right)^{T}  \bm{\alpha} \right]
    \label{eq:pk}
\end{equation}
with $\bm{M}$ a mobility tensor, positive definite. Finally, as in Sec. \ref{sec:theory}, the reversible part of the evolution of the order parameter is $\dot{\psi} = 0$. Adding the dissipative contribution, we arrive at the equation governing the evolution of the phase field,
\begin{equation}
    \dot{\psi} = L \left[ -C_{sh} \frac{\delta \Phi_{sh}}{\delta \psi} - \frac{\delta \Phi_{uq}}{\delta \psi} + C_{\rho} \frac{\delta \Phi_{\rho\psi}}{\delta \psi} \right].
    \label{eq:pf_sd}
\end{equation}
The constant $L > 0$ is a scalar mobility.

The complete set of equations includes mass and momentum conservation, Eqs. (\ref{eq:mass_con}) and (\ref{eq:mom_con}), the simpler kinematic laws valid for small deformations (\ref{eq:kinematics_sd}), and the phenomenological currents in Eqs. (\ref{eq:stress_sd}), (\ref{eq:pk}), and (\ref{eq:pf_sd}).

\section{Discussion and conclusions}

We have reformulated the Phase Field Crystal model to account for the necessary microscopic independence between the phase field, reflecting the symmetry of the phase, and both mass density and elastic distortion. Although these quantities are related in equilibrium through a macroscopic equation of state, they are independent variables in the free energy, and can be independently varied in evaluating the dissipation functional that expresses the Second Law. We have therefore introduced an independent configurational distortion tensor $\bm{P}$ which is a pointwise functional of the phase field $\psi$, but independent of the elastic distortion $\bm{W}$. It captures the local state of distortion of $\psi$, including any topological defects. The latter would be located in regions in which $\curl \bm{P} \neq \bm{0}$, in analogy with the incompatibility condition of the distortion $\curl \bm{W} = - \bm{\alpha}$. In addition, we explicitly include a mass density $\rho$ which is independent of the phase field $\psi$. These considerations assume that the phase field $\psi$ is a non conserved, broken symmetry variable that reflects the symmetry of the system under study, but that is independent of both mass and distortion.

In order to realistically model defect motion in a crystalline phase, choices need to be made in the magnitude of the coupling terms in the free energy linking the phase variable $\psi$ on the one hand, and $\bm{W}$ and $\rho$ on the other. Given a material dependent magnitude of the elastic constant tensor $|\bm{C}|$, we assume that $C_{sh} \sim C_{w} \gg |\bm{C}|$. These conditions ensure fast diffusive relaxation of the phase field to accommodate the existing elastic distortion and topology constraints. As discussed in Sec. \ref{sec:theory}, this is accomplished by having the phase field relax to a local minimum of $C_{sh}\Phi_{sh}+ \frac{ C_{w}}{2} \int d \bm{x}\; \rho |\bm{W}-\bm{P}|^2$, so that the resulting elastic energy and density fluctuations will then decay in their respective time scales. The tensor difference $(\bm{W}-\bm{P})$ plays the role of the compatible strain $\bm{\epsilon}^{\delta}$ of Refs. \cite{re:skaugen18b,re:salvalaglio20}. They are zero in equilibrium, but allow making $\psi$ and $\bm{W}$ independent otherwise.

Allowing the mass density $\rho$ to be independent of the phase field $\psi$ allows for permeation, the independent motion of mass and lattice. In the case of a monocomponent crystalline solid, for example, this dissipative mode has to be understood as vacancy diffusion. Equations (\ref{eq:op_evol}) (or Eq. (\ref{eq:pf_sd}) in the small deformation limit) can be interpreted as permeation equations as their right hand sides equal the normal projection of $\bm{v}-\bm{v}_{\psi}$ along the surface of constant $\psi$, where $\bm{v}_{\psi}$ is the local velocity of such a surface. If $C_{\rho}$ is chosen sufficiently large, then $\rho$ and $\psi$ will locally coincide. However, the ability to separate mass density and phase field is necessary in the treatment of dislocation climb, for example.

The model also naturally incorporates mechanical boundary conditions, either directly applied to the material velocity field $\bm{v}$, or traction involving the stress tensor at the boundary $\bm{T} \bm{\hat{n}}$. The phase field - also with its own natural boundary conditions - will adjust dynamically in the bulk \cite{re:skaugen18}. Solution procedures for the dislocation mechanics part of the problem at small and finite deformations are detailed in \cite{arora2020finite,roy2005finite}; these are non-standard systems taking into account the nonlinear transport of the dislocation density field and the calculation of nonlinear stress fields of dislocation distributions.  The computation of the presented coupled model is material for future work. 

We close by noting that the formulation developed is applicable not only to crystalline solids, but also to other broken symmetry phases such as colloidal, columnar, and smectic phases. 

\section*{Acknowledgments}

JV's research has been supported by the National Science Foundation, Grant No. DMR-1838977.

\bibliographystyle{ieeetr}
\bibliography{pf_plast_arxiv}

\appendix

\section{Notation and definitions}
\label{sec:notation}

We use boldface throughout the paper to denote both vectors and rank two (and four) tensors in three dimensional space. Vector and tensor operations are assumed, including differential calculus. All tensor components are expressed w.r.t the basis of a fixed Rectangular Cartesian coordinate system and all partial derivatives are w.r.t the coordinates of this system. We give here a few explicit definitions in terms of vector and tensor components to avoid possible ambiguity.

If $\bm{A}$ and $\bm{B}$ are two tensors, we define $\bm{A}:\bm{B} = A_{ij}B_{ij}$. Summation over repeated indices is implied. The cross product with a vector $\bm{v}$ is given by $\left( \bm{A} \times \bm{v}\right)_{ij} = \epsilon_{jrs} A_{ir}v_{s}$, where $\epsilon_{jrs}$ is the alternating Levi-Civita tensor. Also, in three dimensions, $\left(\curl \bm{A} \right)_{ir} = \epsilon_{rjk} \partial_{j}A_{ik}$.
%
% which seems to be the transpose to what the math books have.
%
If $\bm{c}$ is an arbitrary vector constant, then  $(\bm{A} \times \bm{v})^{T} \bm{c} = (\bm{A}^{T}c)\times \bm{v}$, $\divergence (\bm{A}) \cdot \bm{c} = \divergence (\bm{A}^{T} \bm{c})$, $(\curl \bm{A})^{T} \bm{c} = \curl (\bm{A}^{T}\bm{c})$. In expressing the Peach-K\"{o}hler force in a suggestive form, we use the symbol $\bm{X}$ to denote the alternating tensor, with $\bm{X}:\bm{A}$ representing a contraction on the last two indices of the alternating tensor.

\end{document}